\documentstyle[11pt]{article}
\begin{document}
\newcommand{\eqn}[1]{(\ref{eq:#1})}
\newcommand{\eq}{\begin{equation}}
\newcommand{\en}{\end{equation}}
\newcommand{\bea}{\begin{eqnarray}}
\newcommand{\eea}{\end{eqnarray}}
\newcommand{\nn}{\nonumber \\ }
\newcommand{\bdm}{\begin{displaymath}}
\newcommand{\edm}{\end{displaymath}}
\newcommand{\A}{\cal{A}_\gamma }
\newcommand{\At}{\tilde{\cal{A}}_{\gamma}}
\newcommand{\Ak}{{\cal A}_{3k}}
\newcommand{\SU}{\widehat{SU(2)}}
\newcommand{\SP}{\widehat{SU(2)^+}}
\newcommand{\SM}{\widehat{SU(2)^-}}
\newcommand{\SSS}{\widehat{SU(3)}_{\tilde{k}^+}}
\newcommand{\br}{\langle}
\newcommand{\kt}{\rangle}
\newcommand{\bra}[1]{\langle {#1}|}
\newcommand{\ket}[1]{|{#1}\rangle}
\newcommand{\lm}{\ell^-}
\newcommand{\lp}{\ell^+}
\newcommand{\la}{\lambda}
\newcommand{\al}{\alpha}
\newcommand{\eps}{\epsilon}
\newcommand{\vl}{\vec{\lambda}}
\newcommand{\pa}{\partial}
\newcommand{\ktp}{\tilde{k}^+}
\newcommand{\ktm}{\tilde{k}^-}

\newcommand{\NP}[1]{Nucl.\ Phys.\ {\bf #1}}
\newcommand{\Prp}[1]{Phys.\ Rep..\ {\bf #1}}
\newcommand{\PL}[1]{Phys.\ Lett.\ {\bf #1}}
\newcommand{\NC}[1]{Nuovo Cim.\ {\bf #1}}
\newcommand{\CMP}[1]{Comm.\ Math.\ Phys.\ {\bf #1}}
\newcommand{\PR}[1]{Phys.\ Rev.\ {\bf #1}}
\newcommand{\PRL}[1]{Phys.\ Rev.\ Lett.\ {\bf #1}}
\newcommand{\MPL}[1]{Mod.\ Phys.\ Lett.\ {\bf #1}}
\newcommand{\IJMP}[1]{Int.\ J.\ Mod.\ Phys.\ {\bf #1}}
\newcommand{\JETP}[1]{Sov.\ Phys.\ JETP {\bf #1}}
\newcommand{\TMP}[1]{Teor.\ Mat.\ Fiz.\ {\bf #1}}
\def\spinst#1#2{{#1\brack#2}}

\begin{titlepage}
\null
\begin{flushright}
DFTT-45/92 \\
September, 1992
\end{flushright}
\vspace{1cm}
\begin{center}
{\Large\bf
The Infrared Limit of
QCD  Effective String \footnote{Lectures given at the XXXII Cracow School of
Theoretical Physics, Zakopane, Poland, 2-12 June 1992}
\par}
\vskip 3em
\lineskip .75em
\normalsize
\begin{tabular}[t]{c}
{\bf F. Gliozzi}\\ \\
{\sl Dipartimento di Fisica Teorica dell'Universit\`a di Torino} \\
{\sl and INFN, Sezione di Torino}\\
{\sl Via P. Giuria 1}, {\sl I-10125 Torino, Italy}
\end{tabular}
\vskip 1.5em
{\bf Abstract}
\end{center} \par
Some model-independent properties of the effective string of gauge
field systems in the confining phase , for very large quark
separations, are described in terms of two-dimensional conformal field
theories. The constraints induced by the gauge theory at the boundaries
of the effective string induce a Coulomb- like term in the interquark
potential which is universal, but different from the one proposed by
L\"uscher. Some universal relations among the string tension, the
thickness of the colour flux tube, the location of the deconfining
temperature and the mass of the lowest glueball state are discussed.
\vskip.3 cm
PACS numbers 11.15.-q
\end{titlepage}
\baselineskip=0.8cm
\renewcommand{\thefootnote}{\arabic{footnote}}
\setcounter{footnote}{0}
The possibility of describing the long-distance dynamics of gauge
theories in the confining phase by an effective string theory is a
fascinating, twenty-years-old conjecture \cite{NO}, which is
resisting against numerous attempts to prove (or disprove) it
\cite{Mi}.
\vskip.3cm
It is based on the very intuitive assumption that the colour flux
connecting a pair of distant quarks is concentrated, in the confining
phase, inside  a thin flux tube, which then generates the linear rising
of the confining potential. Actually this flux tube has been recently
observed in lattice simulations \cite{{DG},{Wo}}.
\vskip.3cm
According to the common lore, this thin flux tube should behave, when
the quarks are pulled very far apart, as a free vibrating string. This
is also supported by the strong coupling expansion of the lattice gauge
theories, which can be formulated as a sum of weighted random surfaces
with quark lines as boundary.
\vskip .3 cm
Unfortunately, the action describing this
effective string in the continuum limit is substantially unknown.
The simplest assumption is that this action is described
 by  the Nambu-Goto string\cite{luscher} which in the conformal gauge
can be described in terms of $D-2$ free bosonic fields associated to the
transverse displacements of the string.

There are, however,  two kinds of difficulties in applying the bosonic
string outside the critical space-time dimension of 26.
\vskip .3 cm
The first is that, depending on the
quantization method, one finds either the breaking of the Lorentz
invariance or the appearance of the conformal Liouville mode,
destroying the simple free field description of the string. If one
insists in  formulating the effective string theory only in terms of
transverse modes, one can replace the Liouville field with a
non-polynomial interaction term \cite{polchi} which looks rather
cumbersome.
\vskip.3cm
The other difficulty for the Nambu-Goto string is its instability
against the formation of pinches for space-time dimension $D>2$ . As a
consequence, the string world sheet takes the shape of  a branched
polymer, which is very different from the behaviour of the
colour flux tube. In order to control this crumpling transition it
has  been considered another kind of string action in which
 a new scale invariant, non-conformal interaction term proportional to
the square of the extrinsic curvature of the world sheet is added to the
Nambu-Goto term
{}~\cite{{helf},{Po}}. This string, sometimes called rigid string, favors
more realistic smooth configurations of the flux tube at short distances.
\vskip.3cm
Note that these modifications, if on one hand transform the Nambu-Goto
action into a consistent theory, on the other
hand make it very difficult to evaluate, even approximately,  physical
observables, which should be the final goal of the effective string
picture of the gauge theories.
\vskip .3 cm
Luckily,  these interaction terms modify the
string theory only at short distance: indeed it  has been shown
that the Lorentz invariance is  asymptotically restored \cite{olesen}
at large distance and that the rigid string does not
modify\cite{bra} the infrared behaviour of the interquark potential
predicted by the Nambu-Goto action. Thus one is led to conclude that
the effective string is asymptotically described\cite{luscher} by a
two-dimensional conformal  field theory formed by $D-2$ massless
free bosonic fields.  \footnote{This is the infrared limit of the
Nambu -Goto action  called in the following the free bosonic string}
\vskip .3 cm
We shall see in this lecture that this description is too drastic an
approximation, because there are constraints  dictated by the gauge
system\cite{cb} which cannot obeyed by the free bosonic string.
There is indeed a simple modification of this picture, consisting in
a suitable compactification of the bosonic fields, which fulfills the
constraints and fit well with the numerical simulations of the gauge
systems in three and four space-time dimensions and with various gauge
groups. We shall see also that this new asymptotic form of the effective
string accounts for the observed finite thickness of the colour flux
tube, gives a good lower bound to the glueball masses and, finally,
suggests a universal value for the transition temperature to the quark
-gluon plasma.

\vskip .3 cm
We start with the rather general assumption that the infrared limit of
the effective string is described by a two-dimensional conformal field
theory (CFT).
Then, the vacuum expectation values of gauge invariant
quantities involving large loops are expressible as the partition
function of this CFT on a Riemann surface with these loops as boundaries.
In particular, in the study of the interquark potential two kinds of
loops are considered: the rectangular Wilson loop and the Polyakov
loop. The rectangular Wilson loop is expressed in terms of the contour
integral of  the gauge field $A_\mu(x)$ along a rectangle  $\rho$
of sides $L$ and $R$ as follows:
\eq
\langle W(R,L)\rangle=\langle tr {\cal P} \exp(\oint_\rho
igA\cdot {\rm d}l)\rangle
\label{wil}
\en
where ${\cal P}$ is the path-ordering and $g$ is the gauge coupling
constant.

The Polyakov loop $P(x)$ can
be defined in a gauge system at finite temperature or, equivalently,
confined in a box with periodic boundary conditions. Then $P(x)$ is
given again as the trace of the path-ordered  exponential of the
contour integral of $A_\mu$ along a line parallel to the periodic
direction and crossing the point $x$.
\vskip .3cm
According to our assumption on the asymptotic form of the effective
string theory, we can write
\eq
\langle W(R,L)\rangle={\rm e}^{-F(R,L)}~~~,
\label{wilson}
\en
where $F(R,L)$ is the free energy of a suitable conformal theory define
on a rectangle.
Similarly, in a $D$-dimensional gauge system at finite temperature
$T=1/L$ or, equivalently, on a box of size
$L\times {\infty}^{D-1}$,
the correlation function of two Polyakov lines $P(x)$ parallel to the
periodic time axis at a distance $R$ is given by
\eq
\langle P(x)P^\dagger(x+R) \rangle =Z_{cyl}=tr{\,\rm e}^{-LH}=
{\rm e}^{-F_{cyl}(R,L)}~~~~,
\label{polya}
\en
\noindent
where $H$ and $F_{cyl}(R,L)$ are now the hamiltonian and the free
energy of the same conformal
theory on the cylinder of height $R$, bounded by the two Polyakov lines
$P$ and $P^\dagger$ of length $L$, which represent the world lines
of a quark and an antiquark, respectively.
\vskip.3cm
In conformal field theory a central role is played by the conformal
anomaly
$c$, which measures the response of the dynamical system to curving
of the surface and
also controls the finite size scaling through a sort of Casimir effect
\cite{cardy}. It can be simply derived by the transformation law of the
holomorphic component of the energy momentum tensor $T(z)$ under the
conformal transformation $z\rightarrow w(z)$:
\eq
T(z)=w'^2T(w)+\frac{c}{12}\{w,z\}~~~,
\label{T}
\en
where the schwarzian derivative $\{w,z\}$ is given by
\eq
\{w,z\}=\frac{w'''}{w'}-\frac{3}{2}\left(\frac{w''}{w'}\right)^2~~~.
\en
A physical state $\ket\psi$ of CFT is  characterized by the Virasoro
constraints
\bea
L_n\ket\psi=0~~,n>0~~~,&L_o\ket\psi=h\ket\psi~~~,\nn
L_n=\frac{1}{2i\pi}\oint\frac{{\rm d}z}{z}z^{n+2}T(z)~~~,&
\eea
where $h$ is the conformal weight of $\ket\psi$ .
Consider now an infinite strip of width $R$ parametrized
by $0\leq\Im m\,z\equiv y\leq R$ . It can be considered as the limit
$L\to {\infty}$ of the cylindric world-sheet of eq.(\ref{polya}), where
\eq
H=\frac{1}{2\pi}\int_0^R T_{oo}{\rm d}{y}=
\frac{1}{2\pi}\int_0^R(T(z)+\bar{ T}(\bar z)){\rm d}y~~~.
\en
The transformation $z\to w=\exp\left(\frac{\pi z}{R}\right)$ maps
conformaly the infinite strip into the upper -half plane
$\Im m \,w\geq0$ with $\{w,z\}=-\frac{\pi^2}{2R^2}$ and
$\bar{T}(\bar w)$ may be taken as the analytic
continuation of $T(w)$ in the lower-half plane\cite{cardy2} .
Then $H$ becomes
\eq
H=\frac{1}{2i\pi}\oint\frac{{\rm d}w}{w}\frac{\pi}{R}\left(w^2T(w)-
\frac{c}{24}\right)=\frac{\pi}{R}\left(L_o-\frac{c}{24}\right)~~.
\label{ha}
\en
The spectrum of the physical states which can propagate along the strip
depends on the boundary conditions (BC) on either side of the strip.
Labelling this pair of conditions with $\alpha$ and $\beta$, and with
$h_{\alpha\beta}$ the conformal weight of the lowest physical state
$\ket{\psi_{\alpha\beta}}$, we have obviously
\def\psia{\ket{\psi_{\alpha\beta}}}
\eq
H\psia=\frac{\pi}{R}\left(h_{\alpha\beta}
-\frac{c}{24}\right)\psia~~~.
\label{ground}
\en
In the limit $L\to {\infty}$ only this state contributes to the free
energy. Then we have
\eq
V(R)=\lim_{L\rightarrow{\infty}}{F_{cyl}(R,L) \over L} =
\sigma R + k_{\alpha \beta}
-({c\over 24}-h_{\alpha \beta}){\pi \over R}+O({1 \over R^2})
\label{frasy2}
\en
\def\h{h_{\alpha \beta}}
where $V(R)$ is the interquark potential, the first term $\sigma R$ is
the bulk contribution which defines the string tension $\sigma$ and
$k_{\alpha\beta}$ is a non-universal constant.
\vskip .3 cm
Similarly, if we take the other limit $R\to{\infty}$ keeping $L$ fixed
in eq.(\ref{polya}), we get the free energy of an infinitely
long cylinder having the asymptotic expansion ~\cite{cardy}
\eq
F_{cyl}(R,L)/R=\sigma L - {{\tilde c }\over 6} {\pi \over L} +
O({1\over {L^2}})~~~,
\label{frasy}
\en
where the combination $\tilde c=c-24 h$ is known as the effective
conformal anomaly~\cite{itz} and  $h$ is the lowest conformal weight
of the states (closed string states) which can propagate along the
cylinder. If the theory is unitary, the lowest state is the vacuum
with $h=0$~.\footnote{The term $-{\pi\tilde{ c}/ 6L}$ can be also
viewed as a universal correction
of the string tension due to the finite temperature $T={1/ L}$, i.e.
{}~\cite{pisarski} $\sigma(T)=\sigma-\pi \tilde cT^2/6+O(T^3)$.}

In the case of the strip (open string) the analysis of the
conformal spectrum is more delicate, because it depends on the choice of
$\alpha$ and $\beta$ in an essential way. In particular, if we
take the same boundary conditions on either side of the strip, it is
possible to show,  as we shall see shortly, that the lowest propagating
state is the vacuum, $i.e.$ $h_{\alpha\alpha}=0$ . Later we shall argue
that in the asymptotic effective string one should have $\h>0$, which
implies $\alpha\neq\beta$. On the contrary, within the Nambu-Goto action
one gets $\h=0$ \cite{luscher}.
\vskip .3 cm
The universal $1/R$ term of eqs. (\ref{frasy2})
and (\ref{frasy}), generated by the non-homogeneous part of the
transformation law of $T(z)$, may also be viewed as a
two-dimensional analog of the Casimir effect, $i.e.$ a universal
contribution to the zero-point energy  due to the finite
size of the dynamical system   . Indeed eq.(\ref{ground})
tells us that the zero-point energy of the CFT on the strip is given by
\eq
E_o=-\frac{\pi(c-24h_{\alpha,\beta})}{24R}~~~.
\label{zenergy}
\en
In a free field theory there is another simple, instructive way to
evaluate $E_o$. For instance, in a free bosonic string of length $R$
with fixed boundary conditions, the physical
degrees of freedom are the $D-2$ transverse normal modes of vibration
described by $D-2$ families of free harmonic oscillators of frequency
$\omega_n=\frac{\pi }{R}n$ . Then $E_o$ is the sum of the zero-point
energy $\frac{1}{2}\hbar\omega_n$ of these oscillators:
\eq
E_o=(D-2)\frac{\hbar}{2}{\sum_{n=1}^{\infty}}'\omega_n=(D-2)
\frac{\hbar\pi}{2R}{\sum_{n=1}^\infty}'n~~~,
\label{zen}
\en
where the apex indicates that the divergent sum has been regularized
in some way. In most cases  regularization introduces a cut-off in
the theory; here it is possible to evaluate unambiguously eq.
(\ref{zen}) without using any cut-off procedure. We only assume that
there is a regularization $f(a)=\sum'(n+a)$ which shares with the
infinite sum $\sum(n+a)$ some of its formal properties, namely that
the (regularized) sum from 1 to ${\infty}$ is equal to the finite sum from
1 to $a$ plus the (regularized) sum from $a+1$ to ${\infty}$ and that the
(regularized) sum of the integers is equal to the sum of the even
integers  plus the sum of the odd integers.

The first condition yields
\eq
{\sum_{n=1}^\infty}'n=\sum_{n=1}^a\,+\,{\sum_{n=a+1}^\infty}'=
\frac{a(a+1)}{2}\,+\,{\sum_{n=1}^\infty}'(n+a)~~,
\label{first}
\en
then
\eq
{\sum_{n=1}^\infty}'(n+a)\,=\,{\sum_{n=1}^\infty}'n\,-\,\frac{a(a+1)}{2}~~~.
\label{fir}
\en
The second and last condition gives
\eq
{\sum_{n=1}^\infty}'n\,=\,2{\sum_{n=1}^\infty}'n\,+\,2{\sum_{n=1}^\infty}'(n-
\frac{1}{2})\,=\,4{\sum_{n=1}^\infty}'n\,+\,\frac{1}{4}~~,
\label{second}
\en
where it has been assumed that eq.(\ref{fir}) is true also for non
integer $a$. As a result we finally get
\footnote{Similarly one may derive, in the same way,
$\sum'n^{2k}=0$ and $\sum'n^{2k-1}=(-1)^k\frac{B_k}{2k}$, where the
$B_k$ are the Bernoulli numbers. These formulas are useful to evaluate
the  Casimir effect (or, equivalently, the Stefan-Boltzmann law) in any
space dimension.}

\eq
{\sum_{n=1}^\infty}'(n+a)\;=\;-\frac{1}{12}\,-\,\frac{a(a+1)}{2}~~~,
\label{reg}
\en
which coincides with the value obtained with the $\zeta$ function
method  \cite{glz,hwk} and other regularizations
\cite{brink,luscher}.

Inserting eq.(\ref{reg}) in eq.(\ref{zen}) we have
\eq
E_o=-(D-2)\hbar\frac{\pi}{24R}~~~.
\label{zebo}
\en
Owing the fixed boundary conditions on either side of the strip, we put
$h_{\alpha\alpha}=0$ in eq.(\ref{zenergy}), then
\eq
c\;=\;D-2~~~,
\label{cbose}
\en
which states the well known fact that each free boson contributes with
$c=1$ to the conformal anomaly.
\vskip.3 cm
It is worth-while to note that the parameter $a$ in eq.(\ref{reg}) is
determined by the choice of $\alpha$ and $\beta$ on the two
sides of the strip. For instance, one might take for $\alpha$  fixed
(or Dirichlet) BC and for $\beta$   free (or Neumann) BC;
this yields $a=-\frac{1}{2}$ and then eqs. (\ref{zenergy}) and
(\ref{reg}) together tell us that the lowest physical state propagating
along the strip has conformal weight $\h=\frac{1}{16}$. This is just an
example of a very general principle of string theory and CFT, stating
that \cite{cardy2}  there is an isomorphism
\eq
\alpha\leftrightarrow \ket{\psi_\alpha}
\en
between the conformally invariant boundary conditions  and the
physical states\footnote{In CFT one works in general  in terms of
primary fields $\phi(z)$ ($i.e.$ BRST- covariant vertex operators in the
language of string theory) which are associated to the irreducible
representations of the Virasoro algebra. The physical state $\ket\psi$
associated to  $\phi(z)$ is created out of the vacuum $\ket0$ simply by
$\ket\psi=\phi(0)\ket0$.}  of the theory: for instance,
the free ends of an open string can consistently interact with a
background field only if this latter describes a physical string state
\cite{ademo}. Another illuminating example is the critical Ising model
\cite{cardy2}; here there are only three boundary conditions which are
invariant with respect to the renormalization group; namely,  the spins
on the boundary can be chosen all up, or all down, or at random
(free BC), and there are only three physical states on the spectrum
of the $c=\frac{1}{2}$ CFT  which describes the  $2D$ Ising model at
criticality.
\vskip .3 cm
The isomorphism mentioned above preserves
the fusion algebra of the CFT, in the sense that the physical states
which can propagate along the
strip with boundary conditions $\alpha\leftrightarrow\ket{\psi_\alpha}$
and $\beta\leftrightarrow\ket{\psi_\beta}$ are of the form\cite{cardy2}
$N^i_{\alpha\tilde\beta}\ket\psi_i$, where the integers $N_{jk}^i$
define the fusion algebra $\phi_j\phi_k=N^i_{jk}\phi_i$ and $\tilde
\beta$ denotes the representation conjugate to $\beta$.
Using this piece of information we can now write the general form of the
partition function on the cylinder defined in eq.(\ref{polya})
\eq
Z_{cyl}(\tau)={\rm e}^{-\sigma RL}N_{\alpha\beta}^i\chi_i(\tau)~~~,
\label{cyl}
\en
where $\tau=\frac{i2R}{L}$ and $\chi_i$ is the Virasoro character of the
representation $i$ propagating along the periodic direction.
\vskip .3 cm
Now we come back to the problem of finding the asymptotic string picture
of the gauge system.
{}From the foregoing discussion, we see that the effective conformal theory
of a gauge system is completely specified not only by the conformal
anomaly $c$ and by the spectrum of the physical states, but also by a
specific choice of the boundary conditions $\alpha$ and $\beta$ on
either side of the strip. Gauge theory poses two important constraints
on the latter two.

\vskip.3cm
Suppose deforming the strip in such a way that the two sides overlap:
if the two boundary conditions $\alpha$, $\beta$ were compatible, we
should get the conformal theory on a cylinder. On the other hand, from
the point of view of the gauge theory, when a quark line overlaps an
antiquark one, the free energy vanishes identically; this is possible
only if $\alpha$ and $\beta$ are incompatible, i.e. the set of states
obeying both $\alpha$ and $\beta$ on
the same side is empty. We can represent this situation symbolically by
\eq
\alpha\cap\beta=\emptyset~~~,
\label{empty}
\en
which implies, in particular
\eq
h_{\alpha\beta}>0~~~,
\label{h}
\en

because the ground state cannot propagate if $\alpha\not=\beta $
{}~\cite{cardy2}.

\vskip.3cm
The other constraint comes simply from the fact that if we perform the
symmetry transformation $A_\mu(x)\to -\, A_\mu(x)$ in the gauge system, the
quark and the antiquark sources are exchanged:
$q\leftrightarrow\bar q$, so there must exist a $Z_2$
automorphism of the CFT which transposes $\alpha$ and $\beta$
\eq
Z_2:\alpha\leftrightarrow\beta\ \ \ F(L,R)\leftrightarrow F(L,R)
\label{z2}
\en

\vskip.3cm
In order to have an effective string which behaves like a colour flux
tube, it is necessary to supplement the theory with some conserved
quantum number, which makes it possible to distinguish the two ends of
the string, as eq.
(\ref{empty}) and (\ref{z2}) require. One possibility, suggested by the
superstring, is to introduce new degrees of freedom, besides the
transverse displacements, which however modify the conformal anomaly
(\ref{cbose}) and are difficult to justify on general grounds.

\vskip.3cm
There is a simpler, more convincing way to modify the bosonic string
without changing $c$, based on the observation that the CFT at fixed
$c\geq1$ are not isolated but depend on a set of parameters
(called the moduli of the CFT). For instance, most of the $c=1$ theories
are described \cite{ginsparg} by a free boson compactified on a circle
of radius $r$.
The free bosonic string corresponds to $r\to \infty$. The spectrum of
physical states is a function of $r$ and it is possible to fix $r$ in
such a way that the BC are fulfilled.

Note also that the real flux tube
connecting a pair of quarks has a small, but not vanishing, thickness.
It could be described classically by  a solution of the type of the
Nielsen-Olesen vortex [1], which has a finite thickness.
Then it is reasonable to expect that the quantum fluctuations around
this classical solution $x^q_i=X_i-X^{cl}_i$ are formed not only by
the local deformations of the string, but also by topological
excitations, characterized by the number of times the string wraps
around the flux tube.  We can implement these winding modes of the
string by assuming the quantum fluctuations  compactified on a circle of
length $2\pi r=L_c$
\cite{cremmer}
\eq
x^q_i\equiv x^q_i+nL_c~~,i=1,\dots D-2~~~.
\label{compact}
\en
It follows that in the functional integral all the string
configurations are equivalent to ones with $\vert x_i^q\vert\leq L_c$,
then they fill a tube of radius
\eq
R_f\simeq\sqrt{D-2}\frac{L_c}{2}
\label{rf}
\en
around the classical solution. We identify this  tube with the colour
flux tube of the gauge theory in the confining phase.

\vskip.3cm
An interesting feature of this one-parameter family  of conformal
theories is that, in many cases\footnote{This happens when
$\nu=\frac{\sqrt{\sigma}R_f}{\sqrt{\pi(D-2)}}$ is rational.}, there is a
free fermion in the spectrum.
 This is a common property of dynamical  systems which can sustain local
and topological modes. A free fermion in the spectrum accounts for
another property of the color flux tubes: they cannot self-overlap
freely but must obey to some constraints which depend on the nature of
the gauge group.
\vskip .3 cm
The two limiting cases are $SU(\infty)$
and $Z_2$ respectively.
In the former case there is no limitation on the overlapping and the
bosonic string could be a good asymptotic description \cite{green}.
On the contrary, in the $Z_2$
case, the flux tube describes in the space-time self-avoiding surfaces.
One may now ask whether the QCD is better described by a free bosonic
string or by a self-avoiding one. Actually it has been shown
\cite{david}  the exact equivalence between a model of random
self-avoiding surfaces embedded in a three-dimensional lattice and a
$O(N)$ lattice gauge theory for any $N$. Moreover this model of
self-avoiding surfaces has a phase transition belonging to
the same universality class of the $Z_2$ gauge model.
A similar construction has also been found for particular $3D$
$U(N)$  lattice gauge theories\cite{neub}.
\vskip.3cm
This suggests assuming that the effective string, at least at large
distances, is described, for any gauge group, by one of  those free
fermion theories mentioned above. The boundary conditions of such a
fermion is a function of $R_f$. We show now
that eq.(\ref{empty}) and (\ref{z2})  fix these boundary
conditions as well as $R_f$.

Consider a free Dirac fermion $\psi(\varsigma,\tau)=\psi^1-i\psi^2$ on
an infinite strip of width $R$ with action
\eq
S=-\frac{i}{2}\int{\rm d}\tau\int_{-\frac{R}{2}}^{\frac{R}{2}}{\rm d}
\varsigma\,\bar{\psi}\not\partial\,\psi~~~~.
\label{fact}
\en
Equations of motion say that
\eq
\psi_\pm^a(\varsigma,\tau)=\psi_\pm^a(\tau\pm\varsigma)~~~~,
\label{pm}
\en
where $\psi_+^a$ and $\psi_-^a$ are the up and down components of the
Majorana fermion $\psi^a$. Because of the finite width, in the variation
of $S$ arises also a boundary term
\eq
\psi_+^a\delta\psi_+^a-\psi_-^a\delta\psi_-^a~~,~~
\varsigma=\pm\frac{R}{2}~,
\label{bul}
\en
which implies $\psi(\varsigma,\tau)\equiv0$ unless one makes the rather
arbitrary assumption $\delta\psi_+=\pm\delta\psi_-$~.

\vskip.3cm
A better way to treat this problem is to add a boundary term $S
_B$~\cite{ademollo} to the bulk action (\ref{fact}) in order to
compensate the contribution (\ref{bul}). The most general conformal
invariant, hermitian term, has the form
\eq
S_B(\varsigma=\frac{R}{2})=\frac{i}{2}{\rm cos}(2\pi\nu)\int{\rm d}\tau
\,\psi_-^a\psi_+^a+\frac{i}{2}{\rm sin}(2\pi\nu)\int{\rm d}\tau\,
\epsilon_{ab}\psi^a_-\psi^b_+~~~,
\label{left}
\en
where $\epsilon_{12}=-\epsilon_{21}=1\,,\,\epsilon_{aa}=0$  and $\nu$
is a not yet specified boundary phase. A similar expression holds for
the other side at $\varsigma=-\frac{R}{2}$ , with another phase $\nu'$,
but it is always possible to redefine the field $\psi$ such that
\eq
S_B(\varsigma=-\frac{R}{2})=\frac{i}{2}\int{\rm d}\tau\,
\psi_+^a\psi_-^a~~~~.
\label{right}
\en
When $\delta S_B$ is combined with eq.(\ref{pm}) and (\ref{bul}) we get
\eq
\psi_\pm(\varsigma+2R,\tau)={\rm e}^{\pm2\pi i\nu}\psi_\pm
(\varsigma,\tau)~~~.
\label{bf}
\en
\vskip.3cm
We come now to the constraints (\ref{empty}) and (\ref{z2}).
Comparing eq.(\ref{left}) and eq.(\ref{right})
 shows that the term proportional to ${\rm cos}(2\pi\nu)$ at $\varsigma=
\frac{R}{2}$ is compatible with that at $\varsigma=-\frac{R}{2}$ ,
therefore the amplitude for a direct open-closed string transition is
forbidden, as required by eq.(\ref{empty}), only if
${\rm cos}(2\pi\nu)=0$, $i. e.$  $\nu=\frac{1}{4}$ or $\nu=\frac{3}{4}$.
\vskip.3cm
To see that also eq.(\ref{z2}) is fulfilled, consider the
reparametrization
$\varsigma\rightarrow-\varsigma$~,which is a symmetry of the bulk
action and transposes the two sides of the strip. We may implement
this symmetry with the following field transformation
\eq
{\cal F}\,:~~\psi^a_+\to\Omega^a_{~b}(\vartheta)\psi^b_-~~,~~
\psi^a_-\to\Omega^a_{~b}(-\vartheta)\psi^b_+~~,
\en
where $\Omega\in SO(2)$ is a rotation of an angle $\vartheta$.
Choosing $\vartheta=\pi\nu$ exchanges also the two boundary terms
(\ref{left}) and (\ref{right}), as eq.(\ref{z2}) requires.
\vskip.3cm
In conclusion, a universal string picture describing the large-distance
behaviour of  gauge theories in the confining phase emerges rather
naturally. It can be formulated simply as a free fermion theory ( one
Dirac fermion for each transverse dimension). Then the normal modes are
now fermi harmonic  oscillators with
\bea
\omega_n=&\frac{\pi}{R}(n-\frac{1}{4})~~,\nn
{\omega'}_n=&\frac{\pi}{R}(n-\frac{3}{4})~~,
\label{freq}
\eea
so we may apply again eqs. (\ref{zen}) and (\ref{reg}) to evaluate the
zero-point energy of this theory:
\eq
E_o^{fermi}=-(D-2)\left[\frac{\hbar}{2}{\sum_{n=1}^{\infty}}'\omega_n\,+\,
\frac{\hbar}{2}{\sum_{n=1}^{\infty}}'{\omega'}_n\right]\,\nn
\en
\eq
{}~=-(D-2)\hbar\frac{\pi}{96R}~~~,
\label{zenf}
\en
which is just one fourth of that of the bosonic string. Comparison with
eq.(\ref{zenergy}) yields
\eq
\h=\frac{D-2}{32}~~~.
\label{hf}
\en
Thus, in order to find the value of the compactification radius we have
to go back to the corresponding bosonic formulation and look for
$c=1$ theories in which the minimal positive conformal weight of the
spectrum is $\frac{1}{32}$. In such a description the primary fields
can be written in the form of vertex operators $:{\rm e}^{ipx^q}:$
with a momentum $p$ given by \cite{cremmer}
\eq
p=mr+\frac{n}{2r}~~,~~m,n\in Z~~~,
\label{mom}
\en
where the adimensional compactification radius $r$ is related to the
scale $L_c$ of eq.(\ref{compact}) as
\eq
r=\frac{\sqrt{\sigma}L_c}{2\sqrt{\pi}}~~~.
\label{ara}
\en
The corresponding conformal weight is $h=\frac{p^2}{2}$ . The theory
of compactified boson has an obvious symmetry $r\to\frac{2}{r}$, known as
duality transformation, which prevents to fix unambiguously the
compactification radius: for each allowed conformal spectrum there are,
in general, two distinct radii, at least, which reproduce it. In the
effective string picture
this duality symmetry is broken at short distance because of the
coupling of the
Liouville mode to the conformal matter \cite{bouka},
so either one of the
compactification radii represents an unstable solution. It is then
reasonable to assume that the physical compactification radius is the
smallest of the possible solutions .
It is easy to see that the theory has a discrete spectrum only if $r^2$
is rational $i.e.~~r^2=\frac{p}{q}$. In such a case the gap $h_{min}$ of
the theory can be written in the general form
\eq
h_{min}\;=\;\frac{5+3(-1)^q}{16\,pq}~~;~~ r^2=\frac{p}{q}~~.
\label{hmin}
\en
Comparison with eq. (\ref{hf}) gives $r=\frac{1}{4}$. We can now
evaluate the thickness of the colour flux tube by translating
this number into physical units. Indeed using eqs. (\ref{ara}) and
(\ref{rf}) yields
\eq
\sqrt{\sigma}R_f=\frac{\sqrt{\pi(D-2)}}{4}~~~.
\label{thick}
\en
Taking for $\sqrt{\sigma}$ the conventional value of 420 $MeV$ and
$D=4$,  we get $R_f\simeq 0.3$ fermi, which
agrees with the value observed in various numerical lattice
simulations \cite{DG,Wo,noi2,noi3}.
\vskip .3cm
The free energy $F(R,L)$ of these CFT's can be written not only in the
asymptotic regions $R/L\ll 1$ or $L/R\ll 1$ described by eqs. (
\ref{frasy2}) and (\ref{frasy}), but for any value of $R$ and $L$.
In particular,
for the rectangular Wilson loop (\ref{wilson}) we get the following
expression
\eq
F_{r=\frac{1}{4}}(R,L)=\sigma RL+ p(R+L) +k+ q(R,L)~~~,
\label{fit}
\en
where the first three terms are the usual  area, perimeter and constant
term ascribed to the classical solution, while  the information on the
CFT is contained in the term $q(R,L)$ produced by the quantum
fluctuations of the string
\eq
q(R,L)=
-(D-2)\log\frac{\vartheta\spinst{1/4}
{1/4}(0|\tau)}{ \eta(\tau)}
\label{q}
\en
where $\tau=iR/L$~, $\vartheta\spinst{\alpha}{\beta}(0|\tau)$ is the
Jacobi theta function with characteristic $\spinst{\alpha}{\beta}$,
and $\eta(\tau)$ is the Dedekind eta function.
Eq.(\ref{fit}) should reproduce the vacuum expectation value of the
Wilson loop only for   values of $R$ and $L$ very large with respect to
the other physical scales entering into the game, so that the
conformal hypothesis is asymptotically true:
\eq
-\log\langle W(R,L)\rangle\simeq F_{r=\frac{1}{4}}(R,L)~~,
{}~~\Lambda R\;,\;\Lambda L\,\gg 1~~,
\label{asy}
\en
where $\Lambda$ is the mass scale of the theory.
\vskip .3 cm
Notice that the string picture which emerges from this description is
exactly that which was proposed some time ago~\cite{noi2} on purely
phenomenological grounds, where the boundary phase $\nu=\frac{1}{4}$
were determined only by a best fit of eq.(\ref{fit}) to the numerical
data on the expectation values of Wilson loops.
This picture as been proven~\cite {noi2,noi3} to be quite good for all
those $3D$ and $4D$ gauge systems where accurate numerical data are
available. In particular, the  values of the string tension $\sigma$
determined by fitting  eq.(\ref{fit}) to the data, show a better
approach to the asymptotic scaling with respect to more conventional
fitting procedures, mostly based on the arbitrary
assumption $q(R,L)\equiv 0$.
\vskip .3 cm
The idea that the asymptotic effective string is described by a
compactified boson allows us also to get approximately the lowest
mass of the glueball spectrum.
This spectrum can be evaluated by studying the exponential decay of the
correlation function of small quark loops at large distance. If $\gamma$
is a small circular loop with center on a point $x$ and $W_x(\gamma)$
is  the  associated Wilson loop operator, we have, asymptotically
\eq
\langle W_x(\gamma)W_{x+L}(\gamma)\rangle\;\sim\;{\rm e}^{-m_GL}~~~,
\label{glueball}
\en
where $m_G$ is the mass of the lowest glueball. In a string picture this
expectation value can be written as the partition function of a CFT on a
surface with the topology of a cylinder of length $L$. However in the
present case , at variance of what happens for the correlation function
of two Polyakov lines, the minimal radius  $R_{\rm min}$ of this
cylinder is not determined by the geometry of the system, rather it
should be generated dynamically. If we assume that  $R_{\rm min}$ is
large  enough to apply CFT formulas, we get from eq.(\ref{frasy})
\eq
\langle W_x(\gamma)W_{x+L}(\gamma)\rangle\;\sim\;\exp\left(-\sigma2\pi
R_{\rm min}L\,+\,\frac{\tilde c\pi}{6}\frac{L}{2\pi R_{\rm min}}
\right)~~,
\label{cft}
\en
where the first term at the exponent is the usual area term and the
second one is due to the universal quantum correction. Comparing eq.(
\ref{glueball}) with eqs. (\ref{cft}) and (\ref{cbose}) yields
\eq
m_G\,\simeq\, 2\pi\sigma R_{\rm min}-\frac{D-2}{12R_{\rm min}}~~~.
\label{mara}
\en
In the bosonic string picture there is no natural lower bound for
$R_{\rm min}$: the minimal area is obtained for $ R_{\rm min}\to0$ ,
where the quantum contribution diverges. On the contrary,  the self-
avoiding string picture we are describing gives obviously
$R_{\rm min}\simeq R_f$, otherwise there is an overlapping of the
colour flux tube. From eqs. (\ref{thick}) and (\ref{mara}) we get
\eq
\frac{m_G}{\sqrt{\sigma}}\,\simeq\,
\sqrt{D-2}\frac{3\pi^2-2}{6\sqrt{\pi}}~~~.
\label{mass}
\en
For $D=4$ we have $\frac{m}{\sqrt{\sigma}}\simeq3.67$, which reproduces
rather accurately the numerical results of the lattice simulations:
indeed for the pure $SU(2)$ gauge theory one finds \cite{baal,mich1}
$\frac{m}{\sqrt{\sigma}}\simeq3.7(2)$, and similarly for pure $SU(3)$
one has \cite{mich2} $\frac{m}{\sqrt{\sigma}}\simeq3.5(2)$.
\vskip .3 cm
The asymptotic string picture we are describing in this lecture allows
us to gain some information also on the phase transition to the quark
gluon plasma \cite{{tc},{saw}}.
Consider indeed a pair of quarks propagating in a gauge medium at a
temperature
$T=1/L$ below the deconfining point $T_c$. According to
eq. (\ref{frasy2}) this system is described by the static potential
\eq
V(R)= \sigma(T)\, R \,-\,c'\frac{\pi}{24R}\,+\,O(1/R^2)~~~,
\label{V}
\en
where for the moment we do not commit ourselves with the value of the
effective conformal anomaly $c'=c-24\h$.
At the deconfining point $T=T_c$ the string tension $\sigma(T_c)$
vanishes and the flux  connecting the two quarks cannot longer be
described by an effective string. As  a consequence,  the long
distance Coulomb-like term $c'\pi/24R$ looses its very justification:
indeed we have seen that this universal behaviour
is produced by the quantum fluctuations of the
string, but now the string has faded away . Consistency requires
vanishing of $c'$  at $T=T_c$, i.e. there must be in the CFT
theory a physical state with a conformal weight
\eq
\h(T_c)=\frac{c}{24}~~~.
\label{hc}
\en
Note that $c'$ , according to the way it has been calculated in
eqs.(\ref{zen}) and (\ref{zenf}), measures the number of local degrees
of freedom of the CFT. Its vanishing tells us that at the deconfining
point the effective string theory has at most a discrete set of degrees
of freedom, $i.e.$ it behaves like a topological conformal field theory
(TCFT).
Actually most TCFT's may be formulated as (twisted)
$N=2$ superconformal theories (SCFT)\cite{witten}. It turns out
\cite{lvw} that
in any $N=2$ SCFT there is a physical state of conformal weight
$h=c/24$ \footnote{It is generated by the spectral flow \cite{schw} of
the Neveu-Schwarz ground state. For further details, see ref. \cite
{lvw}}. Conversely one is led to conjecture
that any CFT with a weight $h=c/24$ is promoted to a $N=2$ SCFT. This
is almost trivially true for $c=1$ and $c=2$, $i.e.$ $D=3$ and $D=4$,
which are the cases we are interested in.
\vskip .3cm
As an example, consider the whole set of $c=1$ CFT's that can be
written in terms of a  compactified boson $x_\perp(z)\,$,
identified later with the field decribing the transverse displacements
of the $D=3$ effective string. The conformal
spectrum is given by eq.(\ref{mom}). If there is a primary field
$:{\rm e}^{p_ox_\perp(z)}:$ with  $h=p_o^2/2=1/24$, then also
$p=6p_o=\pm\sqrt{3}$ belongs to the set of the allowed momenta, of
course. The corresponding primary fields $G(z)=:{\rm e}^{\sqrt{3}x_\perp(z)}$
and $\bar G(z)=:{\rm e}^{-\sqrt{3}x_\perp(z)}$ are precisely the two
supercurrents  which generate the $c=1$ representation \cite{dpz} of
the $N=2$ superalgebra \cite{ademollo}\footnote{The other conserved
currents of this superalgebra are the energy momentum tensor $T(z)$
and the $U(1)$ current $J(z)=\partial x_\perp(z)$}.

On the other hand, using eq.(\ref{hmin}) with $h_{\rm min}=1/24$, we
select four possible compactification radii:

\eq
r_1=\frac{1}{2\sqrt3}\;,\;r_2=\frac{1}{\sqrt3}\;,\;
r_3=\frac{\sqrt3}{2}\;,\;r_4=\sqrt3\;.
\label{rc}
\en
They exactly correspond to the only points where the conformal symmetry
is promoted to a $N=2$ extended supersymmetry. The spectrum of the
primary fields is the same for these four radii, and it is now
described in terms of two quantum numbers: the
conformal weight $h$ and the charge $q$ of the U(1) current.   The
Neveu-Schwarz (NS)  and the Ramond (R) sectors hold each three primary
fields, namely
\bea
NS\,:~~~\{\,(h=0\,,\,q=0)&~,~(h=\frac{1}{6}\,,\,q=\pm\frac{1}{3})~\}\nn
R\,:~~~~\{\,(h=\frac{3}{8}\,,\,q=\frac{1}{2})&~,~
(h=\frac{1}{24}\,,\,q=\pm\frac{1}{6})\,\}~.
\eea
 A possible choice of the boundary conditions $\alpha$ and $\beta$ for
the two sides of the infinite strip obeying to the
constraints (\ref{empty}) and (\ref{z2}) is
\eq
\alpha\leftrightarrow(h=\frac{1}{6}\,,\,q=\frac{1}{3})~~;~~
\beta\leftrightarrow(h=\frac{1}{24}\,,\,q=-\frac{1}{6})~~.
\en
The $Z_2$ automorphism is simply generated by the symmetry between the
two sectors NS and R : $NS\leftrightarrow R
\Rightarrow\alpha\leftrightarrow\beta$.
\vskip.3 cm
Consider now a Wilson loop orthogonal to the imaginary time at a
temperature $T=1/L$ as drawn in fig.1.

Now the field $x_\perp$ describing
the displacements  of the effective string in the direction of
the imaginary time axis is obviously compactified
on a circle of length $L=1/T$, $i.e.$
\eq
x_\perp\equiv x_\perp +L~~~.
\label{tra}
\en
Increasing the temperature of the system reduces the phase space of the
colour flux tube until, for a critical value $T=\Theta$, the inverse
temperature $L$ coincides with the width $2R_f(T)$ of the flux tube:
\eq
2R_f(\Theta)=\frac{1}{\Theta}~~,
\en
where we have taken into account that the thickness of the flux tube may
be a function of the temperature. At this point the flux tube fills the
circle of the imaginary time and is then definitely
trapped in it, so it cannot longer
fluctuate. Then the zero-point energy must vanish and we recover eq.
(\ref{hc}). So we are led to identify $\Theta$ with the deconfinement
temperature $T_c$ and the compactification radii (\ref{rc}) as the
possible values of the size of the flux tube,
according to eq. (\ref{ara}).
\vskip.3cm
\begin{center}
\begin{picture}(380,180)(0,0)
\put(0,80){\line(1,0){200}}
\put(0,0){\line(1,0){200}}
\put(0,0){\line(0,1){30}}
\put(0,40){$\frac{1}{T}$}
\put(0,50){\line(0,1){30}}
\put(200,0){\line(0,1){80}}
\put(0,80){\line(1,1){100}}
\put(0,0){\line(1,1){100}}
\put(200,0){\line(1,1){100}}
\put(200,80){\line(1,1){100}}
\put(100,180){\line(1,0){200}}
\put(100,100){\line(1,0){200}}
\put(100,100){\line(0,1){80}}
\put(300,100){\line(0,1){80}}

\put(60,30){\it Wilson~loop}
\put(50,40){\line(1,0){120}}
\put(50,40){\line(1,1){50}}
\put(170,40){\line(1,1){50}}
\put(100,90){\line(1,0){120}}
\put(51,41){\line(1,0){120}}
\put(52,41){\line(1,1){50}}
\put(172,41){\line(1,1){50}}
\put(101,91){\line(1,0){120}}

\end{picture}
\end{center}
\vskip .3cm
\centerline{\bf figure 1:~\rm A  Wilson loop orthogonal to the imaginary
time}

We are now in a position to evaluate explicitly
$T_c$. Indeed eq.(\ref{rc}) singles out , through eqs.(\ref{tra}) and
(\ref{ara}), four special temperatures.
 It is reasonable to assume that the deconfining temperature corresponds
to the minimal radius $r_1$, while the others correspond to metastable
solutions because, as we have already pointed out, the degeneracy
associated to the eq. (\ref{hmin}) is removed at short distance by the
coupling to the Liouville mode. This gives

\eq
\frac{T_c}{\sqrt\sigma}=\frac{\sqrt{3}}{\sqrt{(D-2)\pi}}~~~.
\label{ng}
\en
The $D$ dependence has been inserted to take into account also the
other  interesting case of $D=4$, which can be treated in the same way.

Indeed at $c=2$ there are two sets of special points in the space of the
CFT's where the symmetry is promoted to an extended $N=2$ supersymmetry.
\vskip .3 cm
A set of these special points corresponds to the direct product of two
$c=1$ $N=2$ theories, where  the string is  described by two identical
compactified bosons $x_i\;,\;i=1,2$ associated to the two transverse
directions. This kind of $N=2$ SCFT is not the correct
description of the asymptotic effective string associated with the
Wilson loop perpendicular to the imaginary time, because in this
configuration the transverse displacements along the two orthogonal
axes are not on the same footing, being only one direction
compactified.
\vskip .3 cm
The other set of $N=2$ SCFT with $c=2$ corresponds
to the $c=2$ element of $N=2$ minimal series, which is described by
two free fields: the first is a compactified bosonic field
$x_\perp(z)$, with a compactification radius related to eq.(\ref{ara})
by $r_c=\sqrt2r_i$, which should describe the string displacements
along the imaginary time; then, using the same arguments as before, we
get for $T_c$ the value stated in eq.(\ref{ng}). The other free field
is a $Z_4$ parafermion $\psi(z)$, which we can associate to the string
displacements along the other direction.
Note that the latter displacements are described by a fermionic field,
according to the effective string theory at $T=0$, where the target
space has no compactified directions.
\vskip .3 cm
Remarkably enough, our determination of $T_c$  coincides with the
value predicted for the Nambu-Goto string~\cite{pisarski}.
Our argument suggests that
this temperature is universal and does not depend on the gauge group.
\vskip .3 cm
In Table I the values of the observables we have determined in
 eqs. (\ref{thick}),(\ref{mass}) and (\ref{ng}) through our
 asymptotic effective string scheme, are compared with the
corresponding data from numerical simulations on lattice
gauge theories with various gauge groups.
\vskip .5 cm
\def\c{\;\cite}
\centerline {\bf Table I: \rm Comparing string values with data from
numerical simulations.}
$$
\vbox{\offinterlineskip
\halign{
\strut\vrule     \hfil $#$ \hfil  &
      \vrule # & \hfil $#$ \hfil  &
      \vrule # & \hfil $#$ \hfil  &
      \vrule # & \hfil $#$ \hfil  &
      \vrule # & \hfil $#$ \hfil  &
      \vrule # & \hfil $#$ \hfil  &
      \vrule # & \hfil $#$ \hfil
      \vrule \cr
\noalign{\hrule}
\noalign{\hrule}
\,{\rm dimension}\,
&&\multispan{5}{~D~=~3~}&&\multispan{5}{~D~=~4~}\cr
\noalign{\hrule}
\noalign{\hrule}
{}~{\rm model}~
&&~{\rm string}~
&&~Z_2~
&&~SU(2)~
&&~{\rm string}~
&&~SU(2)~
&&~SU(3)~\cr
\noalign{\hrule}
\noalign{\hrule}
 ~T_c/\sqrt\sigma ~&&~0.977~&&~1.17(10)\c{tc}&&~0.94(3)\c{fp}
&&~0.691~&&~0.69(2)~~\c{fhk}
&&~0.56(3)~\c{fhk}\cr
\noalign{\hrule}
 ~m_G/\sqrt\sigma~&&~2.596~&&~~&&~~&&~3.671~&&~3.7(2)~~\c{baal}
&&~3.5(2)~~\c{baal}\cr
\noalign{\hrule}
 ~T_c/m_G~&&~0.376~&&~~&&~~&&~0.188~&&~0.180(16)\c{fhk}&&~
0.176(20)\c{fhk}\cr
\noalign{\hrule}
 ~\sqrt\sigma R_f ~&&~0.443~&&~~&&~~&&~0.627~&&~0.4\div0.6~\c{DG}&&~\cr
\noalign{\hrule}
\noalign{\hrule}}}
$$
\noindent
\vskip.3cm
The agreement to the numerical simulations on LGT is in general rather
good and it is conceivable that the few discrepancies are due
essentially to the poor scaling of the current lattice experiments.
It would be interesting to do new numerical simulations in order to
complete the table and to  test the universality of the string
formulas with other  gauge groups.
\vskip .3cm
I would like to thank the organizers of the school for their kind and
warm hospitality, and the partecipants for interesting discussions.

\end{document}